\documentclass[showpacs,twocolumn,amsmath,amssymb,prb]{revtex4}

\usepackage{graphicx,subfigure}
\usepackage{dcolumn}
\usepackage{bm}
\usepackage{color}
\usepackage{ulem}

\newif\iffigure
\figuretrue

\newcommand{\tise}{1{\textit T}-TiSe$_2$}

\begin{document}

\preprint{APS/123-QED}

\title{Revealing the role of electrons and phonons in the ultrafast recovery of charge density wave correlations in 1$T$-TiSe$_2$}

\author{C. Monney$^{1,2}$, M. Puppin$^1$, C.W. Nicholson$^1$, M. Hoesch$^3$, R.T. Chapman$^4$, E. Springate$^4$, H. Berger$^5$, A. Magrez$^5$, C. Cacho$^4$, R. Ernstorfer$^1$, M. Wolf$^1$}


\iffigure
\affiliation{%
$^1$ Fritz-Haber-Insitut der Max-Planck-Gesellschaft, Faradayweg 4-6, 14195 Berlin, Germany,\\
$^2$ Institute of Physics, University of Zurich, Winterthurerstrasse 190, 8057 Zurich, Switzerland,\\
$^3$ Diamond Light Source, Harwell Campus, Didcot OX11 0DE, Oxfordshire, United Kingdom,\\
$^4$ Central Laser Facility, STFC Rutherford Appleton Laboratory, Didcot OX11 0QX, United Kingdom,\\
$^5$ Laboratoire des Nanostructures et Nouveaux Mat\'eriaux Electroniques, Institut de la Mati\`ere Complexe, Ecole Polytechnique F\'ed\'erale de Lausanne, 1015 Lausanne, Switzerland
}%
\fi

\date{\today}
\begin{abstract} 
Using time- and angle-resolved photoemission spectroscopy with selective near- and mid-infrared photon excitations, we investigate the femtosecond dynamics of the charge density wave (CDW) phase in \tise, as well as the dynamics of CDW fluctuations at 240 K. In the CDW phase, we observe the coherent oscillation of the CDW amplitude mode. At 240 K, we single out an ultrafast component in the recovery of the CDW correlations, which we explain as the manifestation of electron-hole correlations. Our momentum-resolved study of femtosecond electron dynamics supports a mechanism for the CDW phase resulting from the cooperation between the interband Coulomb interaction, the mechanism of excitonic insulator phase formation, and electron-phonon coupling.
\end{abstract}

\maketitle

\section{Introduction}

Materials which exhibit a metal-insulator transition concomitant with a structural phase transition naturally raise questions about the driving mechanism of the phase transition, i.e. if it is possible to single out what is the role played by the electronic or lattice degrees of freedom and how they affect the electronic structure of such materials.
In this context, studies of the non-equilibrium dynamics have opened a new dimension in the investigation of correlated materials in recent years. Time- and angle-resolved photoemission spectroscopy (trARPES) has proved to be a major technique, giving access to the momentum-resolved electronic band structure of solids out of equilibrium \cite{SchmittScience,RohwerNature, PetersenPRL, PerfettiPRL, LiuPRB}. In the case of charge density wave (CDW) materials, the most direct approach follows the photoinduced changes of the CDW spectral features, namely the closing of the CDW gap or the disappearance of the CDW backfolded bands directly in the time-domain. Many studies have focussed on the first hundred femtoseconds following the photoexcitation, in order to figure out whether these changes occur faster or not than the lattice motion related to the structural distortion, the timing of which is given by the period of the relevant phonon mode. The recovery dynamics of CDW order have received less attention, however we will show below that they also provide fundamental information on the role of electronic and phonon degrees of freedom.

\iffigure
\begin{figure*}
\begin{center}
\includegraphics[width=14.5 cm]{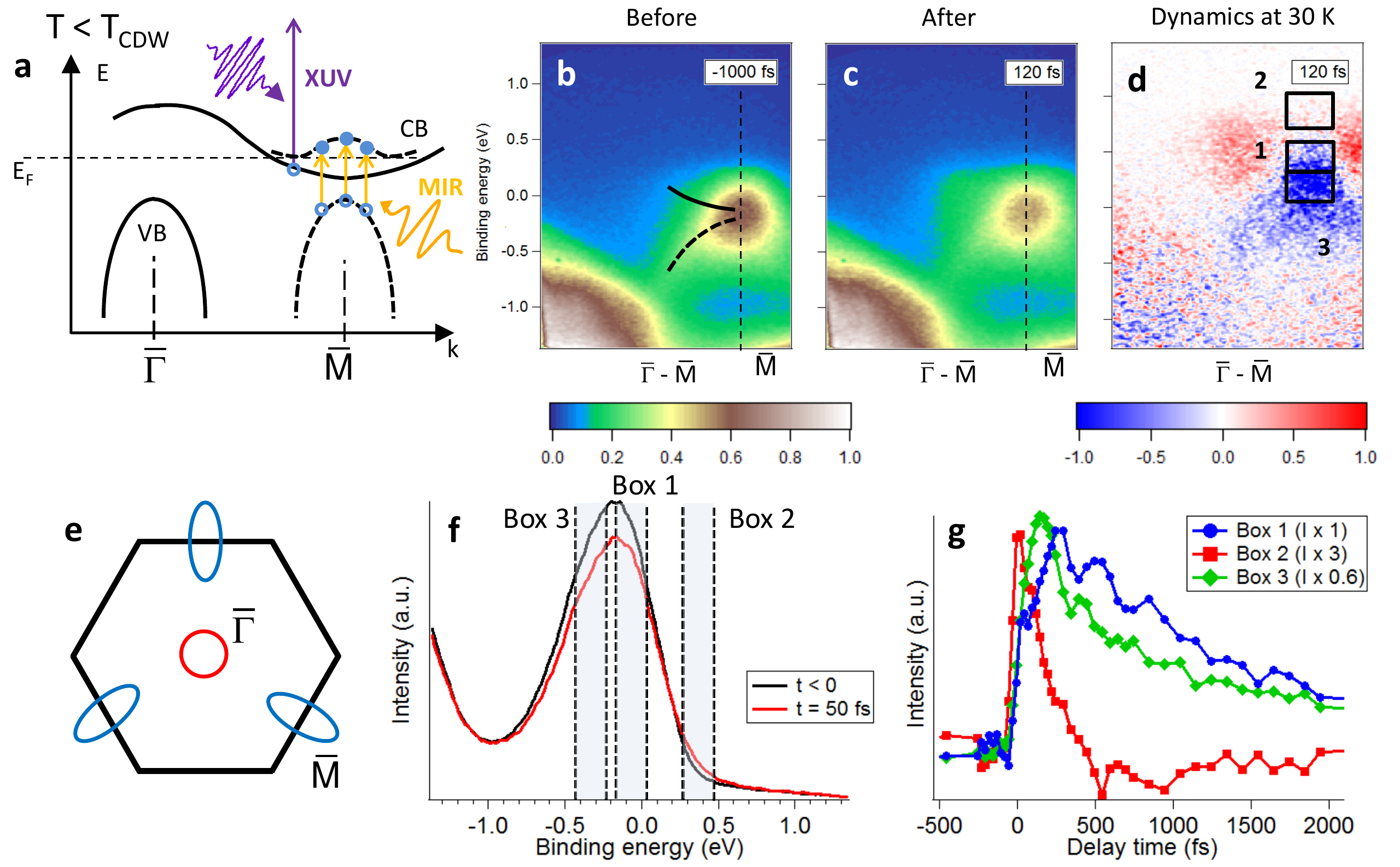}
\end{center}
\caption{\label{fig_1}
(a) Schematic electronic band structure of \tise\  near the Fermi energy in the CDW phase ($T<T_c$), showing the original valence (VB) and conduction (CB) bands (continuous line), and the backfolded VB and CB in the CDW phase (dashed line). (b),(c) Selected photoemisson intensity maps in false colour measured along $\bar{\Gamma}\bar{M}$, near $\bar{M}$, and measured at different time delays, (b) before and (c) after the arrival of the (MIR) 3100 nm (0.4 mJ/cm$^2$) pump pulse, in the CDW phase at 30 K. Guides to the eyes emphasize the relevant bands and the position of $\bar{M}$ is given by the vertical dashed line. (d) Corresponding false colour plots showing the intensity difference between the previous maps. 
(e) Schematic Fermi surface of \tise\ in the normal phase (no CDW) with one hole pocket at $\bar{\Gamma}$ and three symmetry equivalent electron pockets at $\bar{M}$.
(f) EDCs taken at $\bar{M}$ before time 0 (black curve) and at 50 fs (red curve). (f) Absolute value of the photoemission intensity changes integrated in the black boxes of graph (d) (the relative intensities of the curves have been scaled by the factors in the legend for better comparison).
}
\end{figure*}
\fi

CDW materials undergo a low temperature phase transition resulting in a new translational periodicity in the electronic density (the CDW), as well as a periodic lattice distortion (PLD). One of these materials, \tise, undergoes a peculiar second-order CDW transition\cite{DiSalvo} below $T_c=200$ K. The mechanism of its CDW transition has been debated intensively over the last decades, in terms of either an electronic \cite{CercellierPRL, LiPRL} or a structural origin \cite{KiddPRL,CalandraPRL,WeberPRL}. 
Time-resolved techniques seem to give a decisive answer by focussing on the ultrafast melting of the CDW and PLD and disentangling the roles of electrons and lattice on the femtosecond timescale \cite{RohwerNature,HellmannNatComm,MohrPRL}. 
These studies are consistent with a purely electronic mechanism of the CDW transition in \tise, which might be an instability towards an excitonic insulator phase. In such a scenario, electrons and holes from distinct low-energy bands, are bound together in excitons by the interband Coulomb interaction and their macroscopic spontaneous condensation at low temperature leads to a new ground state \cite{KeldyshEI,JeromeEI}.
A recent ultrafast terahertz spectroscopy study\cite{PorerNatureMat} stressed the importance of the lattice degree of freedom, which is shown to stabilize the CDW phase together with the excitonic mechanism.
In parallel, a few theoretical models\cite{vanWezelPRB,ZenkerChiral,KanekoEIphon,ZenkerEIvsLatt} have demonstrated that the CDW (and its concomitant PLD) can be cooperatively stabilized by both the interband Coulomb interaction and the electron-phonon coupling. 
In the light of these recent studies, what is then the role of the excitonic mechanism in \tise ? Does is alone provoke the CDW instability or does it merely assist the lattice in driving the CDW transition?

Here, in order to address these questions, we investigate the femtosecond dynamics of both the recovery of the CDW phase in \tise\ at 30 K, as well as of high-temperature CDW fluctuations at 240 K, which precedes the CDW instability. We exploit the full capabilities of trARPES by looking at the energy- and momentum-resolved recovery dynamics of the CDW correlations. By doing so we unveil the dynamics of both the electron and lattice degrees of freedom. Below $T_c$, in the CDW phase, we observe the coherent oscillations of the CDW amplitude mode for the first time with trARPES. Above $T_c$, we single out a fast component in the recovery of the CDW correlations with a relaxation time $\leq$ 100 fs. We interpret it as the consequence of the ultrafast recovery from the screening of the Coulomb interaction responsible for the electron-hole correlations, which precedes the CDW phase transition. The observation of these two contributions - the coherent oscillations of the CDW amplitude mode at low temperature and the ultrafast electronic recovery at high temperature - supports the idea that both the interband Coulomb interaction and the electron-phonon coupling cooperatively stabilize the CDW phase (and the concomitant PLD) in \tise.

The paper is divided into two main parts: in part \ref{sec_LT}, we study the low temperature (30 K) CDW phase using a mid-infrared excitation. We show how the excitation generates a direct optical transition between the backfolded valence and conduction bands in section \ref{subsec_LT_res}. The resulting coherent oscillation of the amplitude mode is then analyzed and explained as the result of a bonding-antibonding optical transition in section \ref{subsec_LT_phonon}.
Part \ref{sec_RT} shows the results obtained for the CDW fluctuation regime at 240 K using a near-infrared excitation. We demonstrate that the excitation generates a direct transition at the center of the Brillouin zone in section \ref{subsec_RT_res} and subsequently induces a change in the spectral function of the conduction band (section \ref{subsec_RT_SW}). The transient electronic temperature is extracted in section \ref{subsec_RT_Te}. In section \ref{sec_disc}, we discuss all these results and their interpretation within the framework of the CDW phase and fluctations in \tise. We conclude our study in section \ref{sec_conclusion}.

\section{Low temperature CDW phase}
\label{sec_LT}

\subsection{Mid-infrared excitation: CDW resonance}
\label{subsec_LT_res}

Fig. \ref{fig_1} (a) presents the schematic electronic band structure of \tise\ near the Fermi level $E_F$ in the CDW phase. It consists of a valence band (VB) at $\bar{\Gamma}$, the center of the Brillouin zone, and a conduction band (CB) at $\bar{M}$, the border of the Brillouin zone (see Fig. \ref{fig_1}(e) for the corresponding hexagonal Brillouin zone). These two bands are the main components of the low energy band structure at room temperature. Below 200 K, in the CDW phase, a replica of the valence band, the CDW band, appears at $\bar{M}$ as a direct consequence of the new periodicity of the electronic density in the CDW phase, see Fig. \ref{fig_1} (a) (dashed line). In addition, replicas (at symmetry-equivalent $\bar{M}$ point) of the CBs appear above $E_F$\cite{MonneyPRB09,MonneyRIXS,LiPRL}. 

Our pump-probe measurements were obtained with the Artemis time-resolved ARPES endstation at the Central Laser facility of the Rutherford Appleton laboratory \cite{TurcuArtemis}. \tise\ samples were cleaved in a base pressure of $10^{-10}$ mbar and were excited by 100 fs $s$-polarized laser pulses both in the mid-infrared (MIR) region, at 3100 nm (400 meV), and in the near-infrared (NIR) region, at 1400 nm (890 meV). The sample was probed by 30 fs $p$-polarized laser pulses with 21 eV photons. The total energy resolution was about 300 meV. 

In Fig. \ref{fig_1}, we show trARPES data measured in the CDW phase at 30 K when exciting the sample with MIR pump pulses at 3100 nm with a fluence of 0.4 mJ/cm$^2$. At this wavelength, electrons are excited resonantly at $\bar{M}$ between the backfolded valence band and the backfolded CB (see the orange arrows on Fig. \ref{fig_1} (a)).
Fig. \ref{fig_1} (b) and (c) show trARPES intensity plots taken before the arrival of the pump pulse, $\Delta t=-1000$ fs (i.e. at thermal equilibrium) and after the arrival of the pump pulse, $\Delta t=120$ fs, at which time the pump-induced effects are the strongest. At 30 K, \tise\ is in the CDW phase, meaning that a well-developed CDW band is measured at $\bar{M}$ below the CB. However, due to the limited energy resolution of our experiment, these two bands cannot be resolved Fig. \ref{fig_1} (b) and (c) and the CB can hardly be distinguished \footnote{At 30 K, the CDW band at $\bar{M}$ is very intense (see Supplementary Materials\cite{SuppMat}) and dominates over the CB, so that it dominates also the spectral weight distribution when convolved by the experimental energy resolution.}. Nevertheless, comparing the spectra measured at $\Delta t=120$ fs with those at equilibrium, one distinguishes clear differences, which are more obvious in the corresponding difference image in red/blue false colour plot. Such a difference intensity plot is shown in Fig. \ref{fig_1} (d), for $\Delta t=120$ fs, giving a clear view of the dynamics in our trARPES data. It emphasizes the main dynamical changes, namely the disappearance of the CDW band (in blue) and the accumulation of excited electrons in the CB (in red). 

The resonant MIR excitation (400 meV) taking place at $\bar{M}$ also transfers population from the top of the CDW band into the unoccupied states. This can be seen directly by looking at the time-dependent photoemission intensity integrated in the different energy versus momentum regions depicted by boxes in Fig. \ref{fig_1} (d). The energy position of such boxes is also indicated in Fig. \ref{fig_1} (f), where we plot energy distribution curves (EDCs) taken before (black) and 50 fs after photoexcitation (red).
Here the photoinduced changes below the Fermi level $E_F$ are mostly due to the change of the spectral function of the CDW band. However, part of it is also due to the resonant optical transition taking place at $\bar{M}$ with the MIR pump pulse, as schematically displayed in Fig. 1(a). This can be observed in the transient photoemission intensity curves shown in Fig. \ref{fig_1} (g), integrated from the corresponding boxes in graph (d). 
From box 1 (blue curve), taken near $E_F$, an oscillation is observed and will be discussed in the next section. At time 0, an additional sharp peak can be distinguished, which is coming from the depopulation of the CDW band due to the pump pulse. There is correspondingly a peak at about 0.4 eV higher in energy, in the unoccupied states. This can be seen in the time-resolved intensity of box 2 (red curve) where we observe a transient population  of a (backfolded) conduction band at $\bar{M}$.
This indicates that at time 0, electrons are transferred from occupied to unoccupied states at $\bar{M}$, i.e. that an optical transition takes place at $\bar{M}$.
Finally, the green curve shows the intensity integrated in box 3, which is the region where the reduction of the photoemission intensity is the strongest. However, we avoid this region in our further analysis at low temperature, as the nature of the transient signal here is not clear. Indeed, due to the significant energy resolution of our experiment (about 300 meV), the signal coming from the CDW band cannot be well separated from the one from the conduction band at $\bar{M}$ exactly (see also the Supplementary Materials for a high-energy resolution spectrum taken at $\bar{M}$ exactly\cite{SuppMat}).
\begin{figure}
\begin{center}
\includegraphics[width=8.5cm]{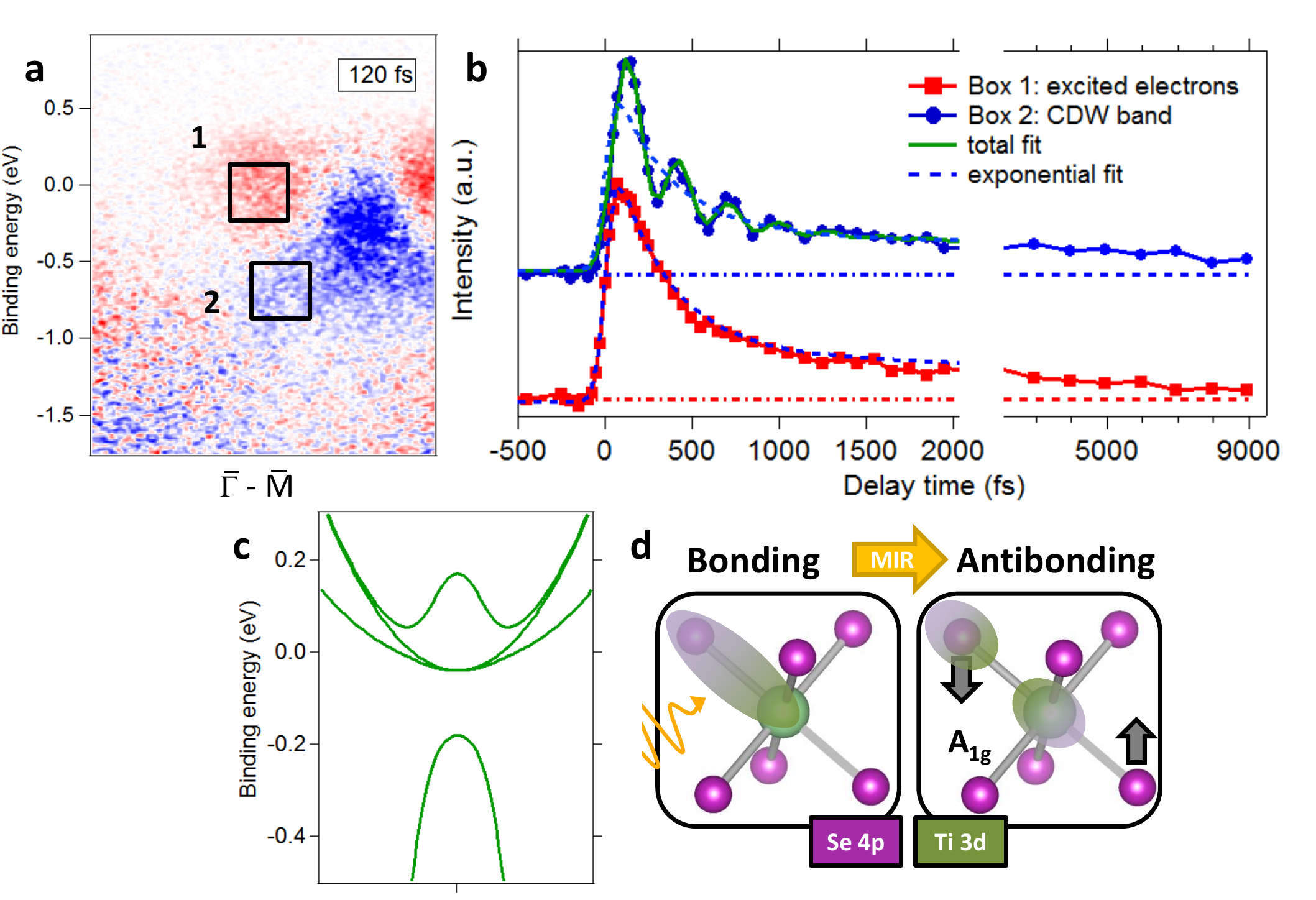}
\end{center}
\caption{\label{fig_2}
(a) Photoemisson intensity map in false color showing the photoemission intensity difference map taken at 30 K for a MIR pump pulses (same as in Fig. \ref{fig_1} (d)).
(b) Absolute value of the normalized photoemission intensity changes integrated in the black boxes of graph (a), together with the fit of the coherent phonon oscillation (total fit, green curve) and the exponential part of the fit only (dashed blue curve).
(c) Schematic electronic structure of \tise\ in the CDW phase at $\bar{M}$.
(d) Schematic of the displacements involved by the $A_{1g}$ coherent phonon excitation by the MIR pump pulse for a TiSe$_6$ octahedron.
}
\end{figure}

\subsection{Coherent phonon oscillation of the CDW mode}
\label{subsec_LT_phonon}

We now look in more detail at the recovery dynamics of the spectral features associated with the excited electrons and with the CDW. The (normalized) time-dependent photoemission intensity of these features is shown in Fig. \ref{fig_2} (b) . The melting dynamics of the CDW band has been already extensively studied and shown to be faster in time for larger pump fluences, an effect assigned to the screening by the transient generation of free charge carriers and therefore consistent with a purely electronic mechanism. \cite{RohwerNature,HellmannNatComm}.
Here, these transient intensities have been integrated from the regions shown by boxes in Fig. \ref{fig_2} (a), corresponding to the excited electrons in the CB (red curve with square markers) and the bottom of the backfolded VB (blue curve with discs). 
They both show an ultrafast response which recovers mostly within 2 ps followed by a slower recovery which is not over at 9 ps. 
Most interesting are the dynamics within 2 ps. We emphasize that for states at binding energies larger than 400 meV (the pump photon energy) there are no unoccupied final states available for direct photoexcitation. The dynamics observed in the CDW band is therefore due to a photoinduced change of its spectral function. The photoemission intensity integrated at the bottom of the backfolded VB displays a clear oscillation on top of an exponential decay, up to about 1 ps. The oscillation can be well described by a cosine with a period of about 290 fs, which correspond to 3.4 THz, and  a phase of about $(0.0\pm0.2)\pi$. 

These numbers are obtained by fitting the transient photoemission intensity of the coherent phonon oscillation using the following function
\begin{eqnarray}
I(t)&=&A_0\exp(-t/\tau_0)+A_1\exp(-t/\tau_1)\nonumber\\&+&A_{ph}\cos(\omega_{ph}t+\phi_{ph})\exp(-t/\tau_{ph}).\nonumber
\end{eqnarray}
This represents a sum of two incoherent exponential components, caracterized by two amplitudes, $A_0$ and $A_1$, with two relaxation times, $\tau_0$ and $\tau_1$, together with an oscillatory component, characterized by an amplitude, $A_{ph}$, a frequency, $\omega_{ph}$, a phase, $\phi_{ph}$, and damped by an exponential of damping time $\tau_{ph}$. The two incoherent exponential components are necessary to achieve the best fit.
This function is then convolved with a gaussian having the experimental time resolution as a full width at half maximum (100 fs).
The following values give the best fit. For the relative amplitudes, $A_0=1$, $A_1=0.23$ and $A_{ph}=0.8$. For the relaxation times, $\tau_0=320$ fs, $\tau_1=7100$ fs and $\tau_{ph}=362$ fs. The phonon period is $2\pi/\omega_{ph}=287$ fs and the phase is 0.

This mode has already been observed in time-resolved reflectivity data from \tise\ \cite{MohrPRL} and was attributed to the $A_{1g}$ amplitude phonon mode of the CDW \cite{SnowPRL,HolyPRB}, corresponding to a breathing mode of a TiSe$_6$ octahedron, as depicted in Fig. \ref{fig_2} (d) \cite{HolyPRB}. This oscillation is highly localized in momentum space and is absent from the dynamics of the excited electrons.
Such a coherent phonon oscillation is caused by the excitation of coherent ion motions by the pump laser pulse. The 0 phase of a cosine-like oscillation indicates that it is a displacive excitation, meaning that the potential energy surface of the corresponding ions, which is suddenly modified by the electronic excitation, triggers the coherent phonon oscillation.
In Fig. \ref{fig_2} (b), the dynamics of the excited electrons and that of the exponential decay of the CDW band are very similar (compare the exponential decay fit of the CDW band intensity (dashed blue curve) with the one of the hot electrons in Fig. \ref{fig_2} (b)), supporting the idea that the transient change of the potential energy surface is modified by the excited electron distribution.
\begin{figure}
\begin{center}
\includegraphics[width=8.5cm]{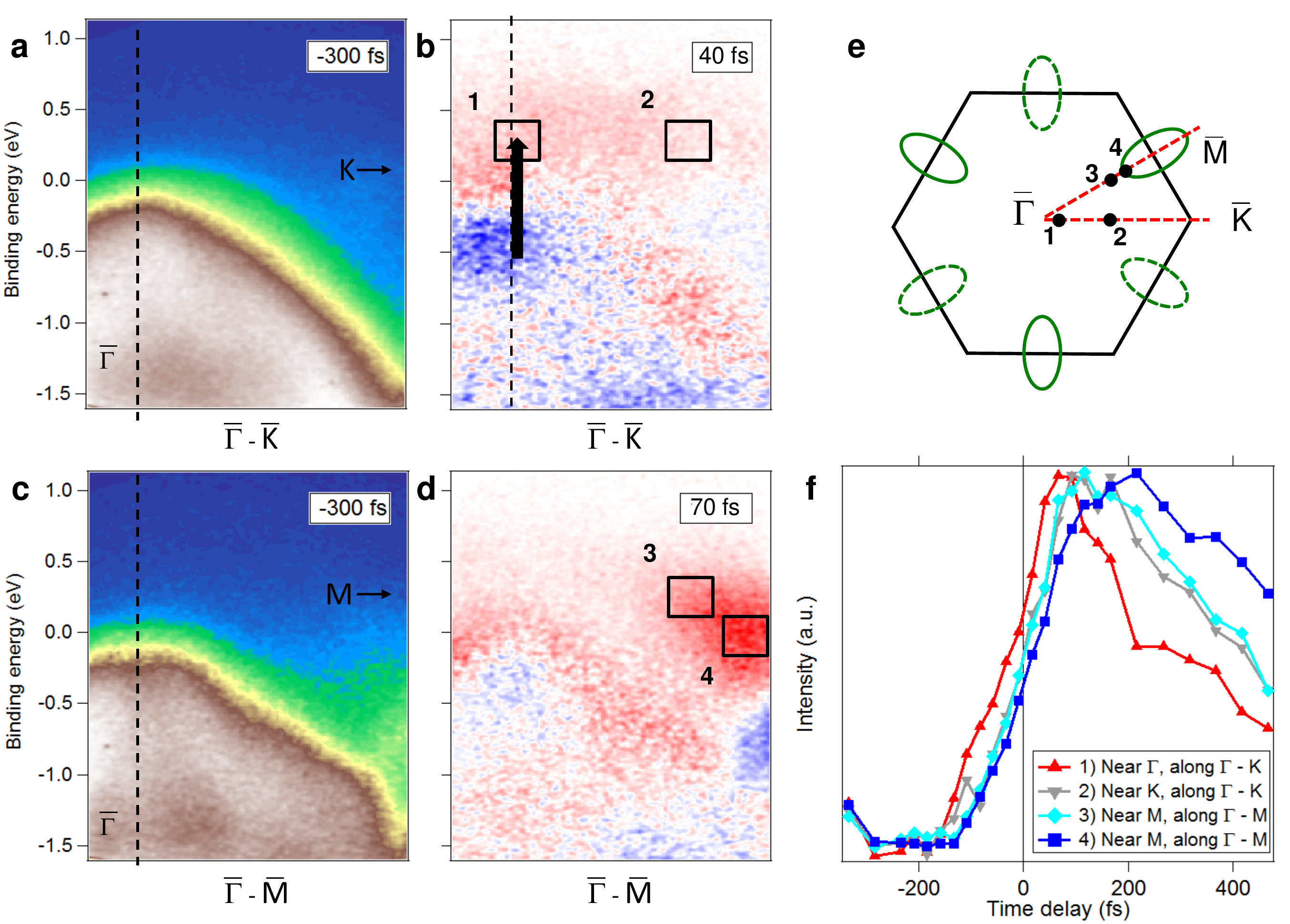}
\end{center}
\caption{\label{fig_3}
(a) Photoemisson intensity map in false color showing the VB along $\bar{\Gamma}\bar{K}$, near $\bar{\Gamma}$, measured at 30 K before time 0 with a 1300 nm pump photon wavelength. (b) Corresponding false color plots showing the intensity difference induced by the pump pulse (positive and negative intensity changes in red and in blue, respectively) near $\bar{\Gamma}$. Electrons and holes are excited resonantly at time 0 by the pump pulse, the energy of which is shown by the black arrow. (c,d) Corresponding intensity maps along $\bar{\Gamma}\bar{M}$, near $\bar{\Gamma}$, where part of the conduction band and the CDW band can be seen on the right side of the graphs. (e) Hexagonal Brillouin zone of \tise, with the two directions relevant to graphs (a,b,c,d). (f) Photoemission intensity traces as a function ot time delay integrated in the corresponding numbered black boxes on graphs (b,d). The curves are normalized to their maximum.  
}
\end{figure}
\iffigure
\begin{figure*}
\begin{center}
\includegraphics[width=14.5 cm]{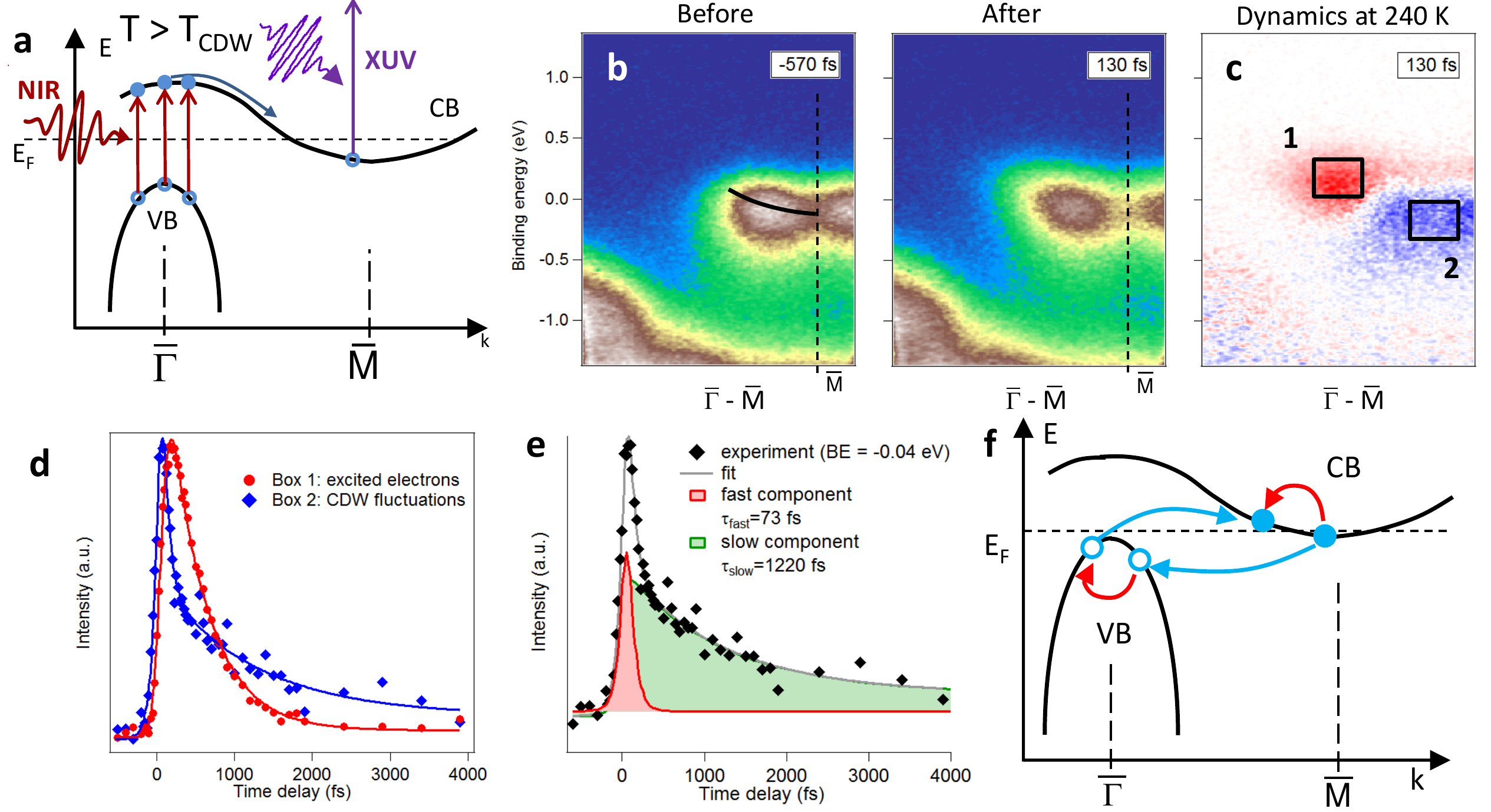}
\end{center}
\caption{\label{fig_4}
(a) Schematic electronic band structure of \tise\  near the Fermi energy in the normal phase ($T>T_c$), showing the original valence (VB) and conduction (CB) bands (continuous line).
(b) Selected photoemisson intensity maps measured along $\bar{\Gamma}\bar{M}$, near $\bar{M}$, and measured at different time delays, before (left) and after (right) the arrival of the (NIR) 1300 nm (0.5 mJ/cm$^2$) pump pulse, at 240 K (above $T_c$). Guides to the eyes emphasize the relevant CB and the position of $\bar{M}$ is given by the vertical dashed line. (c) Corresponding false colour plots showing the intensity difference between the previous maps. (d) Absolute value of the normalized photoemission intensity changes integrated in the black boxes of graph (c), with fits (lines) as guide to the eyes.
(e) Transient photoemission intensity at the bottom of the CB at $\bar{M}$ (black box in the blue region of graph (b), at a binding energy (BE) of -0.04 eV), together with the fitted slow (green) components (see text). 
(f) Schematics of electron-electron scattering between the VB and CB (blue arrow) and equivalently of electron-hole scattering (red arrows).
}
\end{figure*}
\fi
In the CDW phase, it has been shown that the electronic structure of \tise\ can be well described by a mean-field Hamiltonian describing one hole-like band at $\bar{\Gamma}$ coupled to three symmetry equivalent electron-like bands near $\bar{M}$ (see the corresponding schematical Fermi surface in Fig. \ref{fig_1} (e)) \cite{MonneyPRB09}. At $\bar{M}$, the three conduction bands are degenerate (without any coupling) and the effective Hamiltonian has the form
\begin{equation}
H=
\begin{pmatrix}
\varepsilon_a(\vec{k}) & \Delta & \Delta & \Delta \\
\Delta & \varepsilon_{b1}(\vec{k}) &0 & 0 \\
\Delta & 0 & \varepsilon_{b2}(\vec{k}) & 0 \\
\Delta & 0 & 0 & \varepsilon_{b3}(\vec{k})
\end{pmatrix},\nonumber
\end{equation}
where the order parameter $\Delta$ describes the strength of the coupling between the single valence band $\varepsilon_a(\vec{k})$ and the three symmetry equivalent conduction bands $\varepsilon_{bi}(\vec{k})$ ($i=1,2,3$), in the basis of the wave functions for the corresponding electronic states, namely $\{ \psi_a(\vec{k}),\psi_{b1}(\vec{k}),\psi_{b2}(\vec{k}),\psi_{b3}(\vec{k}) \}$. Once the coupling between the valence and conduction bands takes a finite value, the valence band gets backfolded to $\bar{M}$ and the conduction bands get backfolded to $\bar{\Gamma}$ and other $\bar{M}$ points. The new band dispersions along the high-symmetry direction $\bar{\Gamma}\bar{M}$ are shown in Fig. \ref{fig_2}(c) (for realistic dispersions fitted from the experiment\cite{MonneyPRB09}).
For the extrema of the valence and conduction bands, we use the values $\varepsilon_a=30$ meV, $\varepsilon_{bi}=-40$ meV at $\bar{\Gamma}$ and $\bar{M}$, respectively. For the order parameter, we use $\Delta=100$ meV, which fits well with the experiment at low temperature. Diagonalizing the Hamiltonian $H$ leads to the following eigenvalues at $\bar{M}$
\begin{equation}
\varepsilon_1=172\;\text{meV},\quad\varepsilon_2=-182\;\text{meV},\quad\varepsilon_3=\varepsilon_4=-40\;\text{meV}\nonumber.
\end{equation}
The values $\varepsilon_1$ and $\varepsilon_2$ correspond to the backfolded valence and backfolded conduction bands (at $\bar{M}$). The mid-infrared pump photon energy of 400 meV matches approximatively the resonance between these two bands, which can be defined as the CDW gap in \tise. The eigenvectors corresponding to these eigenvalues are
\begin{eqnarray}
\text{for}\qquad\varepsilon_1&=&172\;\text{meV},\nonumber\\
\qquad\psi_1&=&-0.77\psi_a-0.37(\psi_{b1}+\psi_{b2}+\psi_{b3})\nonumber,\\
\text{for}\qquad\varepsilon_2&=&-182\;\text{meV},\nonumber\\
\qquad\psi_2&=&-0.63\psi_a+0.45(\psi_{b1}+\psi_{b2}+\psi_{b3})\nonumber.
\end{eqnarray}
As a consequence, $\varepsilon_1$ corresponds to the antibonding state and $\varepsilon_2$ to the bonding state.

In \tise, the VB at $\bar{\Gamma}$ is mostly of Se $4p$ character and the CB at $\bar{M}$ of Ti $3d$ character. As shown above, in the CDW phase, these bands will effectively hybridize due to their coupling and have a mixed character. We have just demonstrated that the states at the top of backfolded VB band have a bonding symmetry, while those at the bottom of the backfolded CB have an antibonding symmetry. This suggests how the resonant MIR excitation triggers a displacive $A_{1g}$ CDW amplitude mode by exciting electrons from bonding to antibonding states in \tise\ (see Fig. \ref{fig_2}(d)).

Summarizing, in the low temperature phase of \tise , the resonant MIR excitation triggers a coherent lattice response of the CDW amplitude mode, observable in the recovery of the backfolded valence band of the material. This differs from the case of higher pump photon energies \cite{RohwerNature} for which no coherent oscillations are observed. We explain it here by the fact that this resonant optical transition occurs between bonding and antibonding states (at $\bar{M}$) of the TiSe$_6$ octahedron acting efficiently on the ionic position relevant to the CDW distortion and stimulating this way a displacive excitation.

\section{High temperature phase}
\label{sec_RT}

\subsection{Near-infrared pumping: resonance at $\bar{\Gamma}$}
\label{subsec_RT_res}

We now turn to the high temperature trARPES data. The phase transition in \tise\ is continuous (second-order) \cite{DiSalvo} and photoemission data\cite{MonneyPRB12} have shown that short-range CDW fluctuations persist well above $T_c=200$ K. 
Such fluctuations are known to lead to spectral features anticipating those of the low temperature CDW phase, e.g. with a CDW pseudo-gap\cite{GruenerBook,ZrTe3Flucts}. In the case of \tise,
no CDW band is observed, but a broad CB at $\bar{M}$ (see Supplementary Materials for a high-energy resolution ARPES spectrum\cite{SuppMat}), the width of which results from the short-range CDW fluctuations. We have therefore also measured \tise\ at 240 K, above $T_c$, to investigate the dynamics of these spectral features. 
At 240 K, due to the disappearance of the sharp CDW bands at high binding energies, a resonant MIR excitation cannot be used (accordingly, the CDW peak in the optical conductivity vanishes \cite{LiPRL}). We therefore used a shorter wavelength, 1300 nm, in the NIR regime.

The trARPES data of Fig. \ref{fig_3} measured near $\bar{\Gamma}$, along $\bar{\Gamma}\bar{K}$ (see graph (a)), demonstrate that the NIR pulses excite electrons by an optical transition taking place at around $\bar{\Gamma}$, as illustrated in Fig. \ref{fig_4} (a). This direction in the Brillouin zone is depicted in Fig. \ref{fig_3} (e).  Shortly after time 0 (at $\Delta t=40$ fs), one sees a clear depletion of intensity at the top of the valence band (VB) at $\bar{\Gamma}$ and accordingly strong intensity above the VB, in a conduction band (CB), which is slightly less than 1 eV above (see graph (b)) the top of the VB. The black arrow has a length of 890 meV, demonstrating this electronic resonance with the pump pulse. At later times, the excited hot electrons are scattered away from $\bar{\Gamma}$ (see graphs (c) and (d)) and accumulate into the CB  at $\bar{M}$. The time delay difference between these different events is shown in Fig. \ref{fig_3} (f).
This behavior of the momentum path of the hot electrons in the unoccupied states is similar to what has been already observed by Rohde \textit{et al.} \cite{RohdeEPJB} for a much larger pump photon energy.

\subsection{Spectral function changes at $\bar{M}$}
\label{subsec_RT_SW}

Photoemission intensity maps taken at 240 K with NIR excitation are shown in Fig. \ref{fig_4} (b) at negative ($\Delta t=-570$ fs) and positive (at $\Delta t=130$ fs) time delays. At 240 K, i.e. above the critical temperature of the CDW phase, there is not clear CDW band at high binding energies anymore. Our trARPES data have been acquired with a pump photon wavelength of 1300 nm resonant with an optical transition at $\bar{\Gamma}$ (see previous section).
The difference image in red/blue false colour plot, Fig. \ref{fig_4} (c), shows the accumulation of excited electrons in the CB (in red) and a decrease of photoemission intensity at the bottom of the CB (in blue)  
The transient photoemission intensities of the excited electrons and of the bottom of the CB are displayed in Fig. \ref{fig_4} (d). The corresponding time-dependent photoemission intensities integrated in the corresponding boxes in this graph are shown in Fig. \ref{fig_4} (e). Interestingly the decrease of intensity at the bottom of the CB at $\bar{M}$ (in blue) displays a dynamics with 2 timescales. Its recovery dynamics are therefore fitted with a bi-exponential function (convoluted with a gaussian for experimental time resolution), involving a fast and a slow component, with decay constants $\tau\cong 70$ fs and $\tau\cong 1220$ fs, respectively.
We argue below that this decrease of intensity is due to a photoinduced change of the spectral function. 

\begin{figure}
\begin{center}
\includegraphics[width=8.5cm]{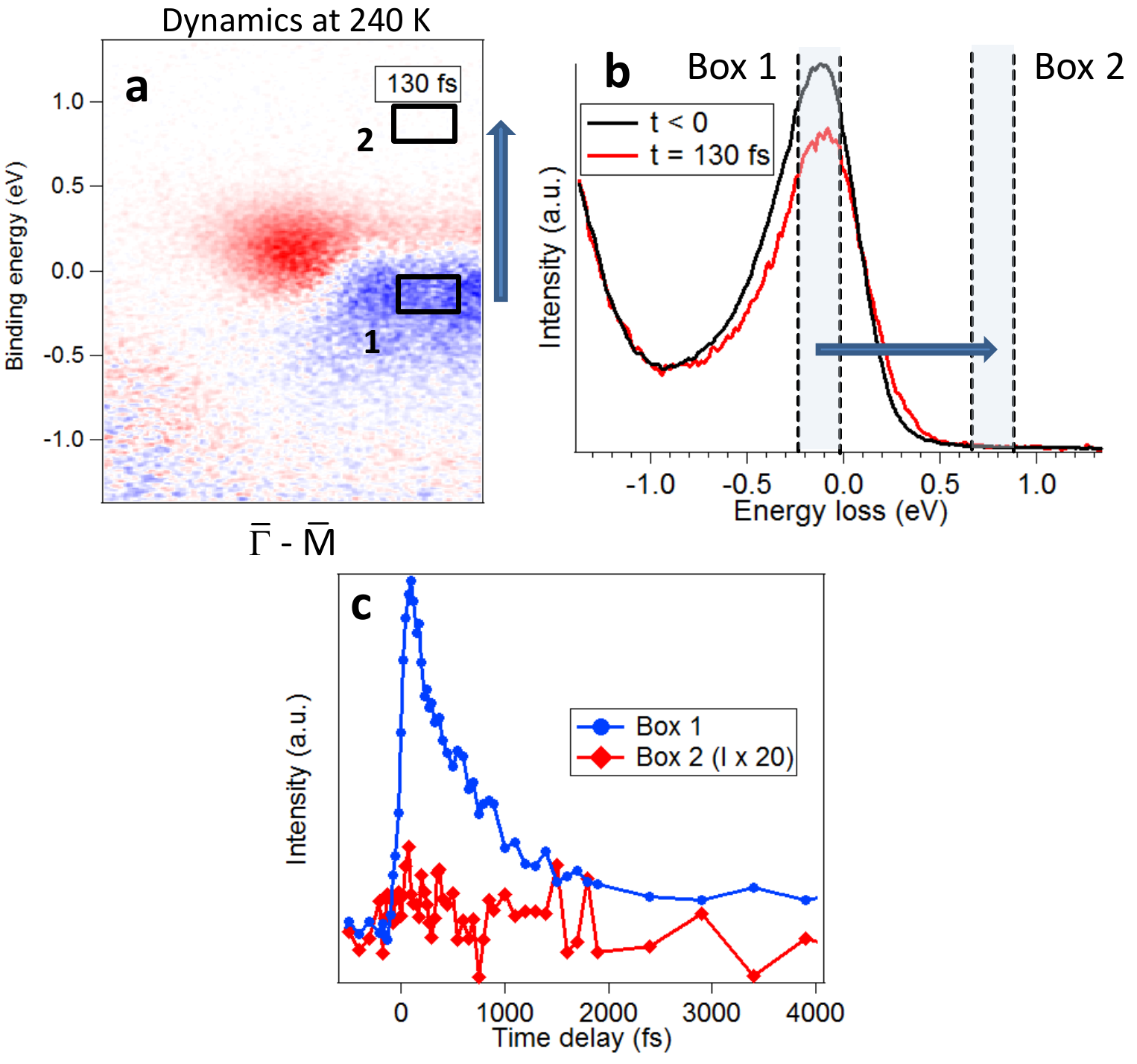}
\end{center}
\caption{\label{fig_6}(a) Photoemisson intensity map in false color showing the photoemission intensity difference map taken at 240 K (same as in Fig. \ref{fig_4}(c)). The blue arrows indicate the size of the pump photon energy (0.95 eV, 1300 nm) in comparison to the energy axis. (b) EDCs taken at $\bar{M}$ before time 0 (black curve) and at 130 fs (red curve). (c) Absolute value of the normalized photoemission intensity changes integrated in the black boxes of graph (a).
}
\end{figure}

In Fig. \ref{fig_6}, we demonstrate that there is no resonant excitation taking place at $\bar{M}$ at 240 K with a pump photon wavelength of 1300 nm (0.95 eV), in the NIR regime. Graph (b) in Fig. \ref{fig_6} displays EDCs taken at $\bar{M}$, before time 0 (black) and at 130 fs (red). Graph (c) shows the transient photoemission intensity integrated in the boxes 1 and 2 in graph (a). Box 1 is situated in a region where the transient suppression of the photoemission signal is the largest and box 2 is situated 0.95 eV higher in energy, in the unoccupied states. The red curve from box 2 shows clearly that there is hardly any transient signal in this region, proving that there is no resonant excitation taking place at $\bar{M}$ at 240 K for 1300 nm pump photon wavelength (at variance with the case of MIR pump photons at 30 K, see Fig. \ref{fig_1}). We can therefore attribute the photemission intensity changes observed below $E_F$ to a photoinduced change of the spectral function of the conduction band, which we will explain further below in our discussion of the results.

\subsection{Extracting the electronic temperature at $\bar{M}$}
\label{subsec_RT_Te}

\begin{figure}
\begin{center}
\includegraphics[width=8.5cm]{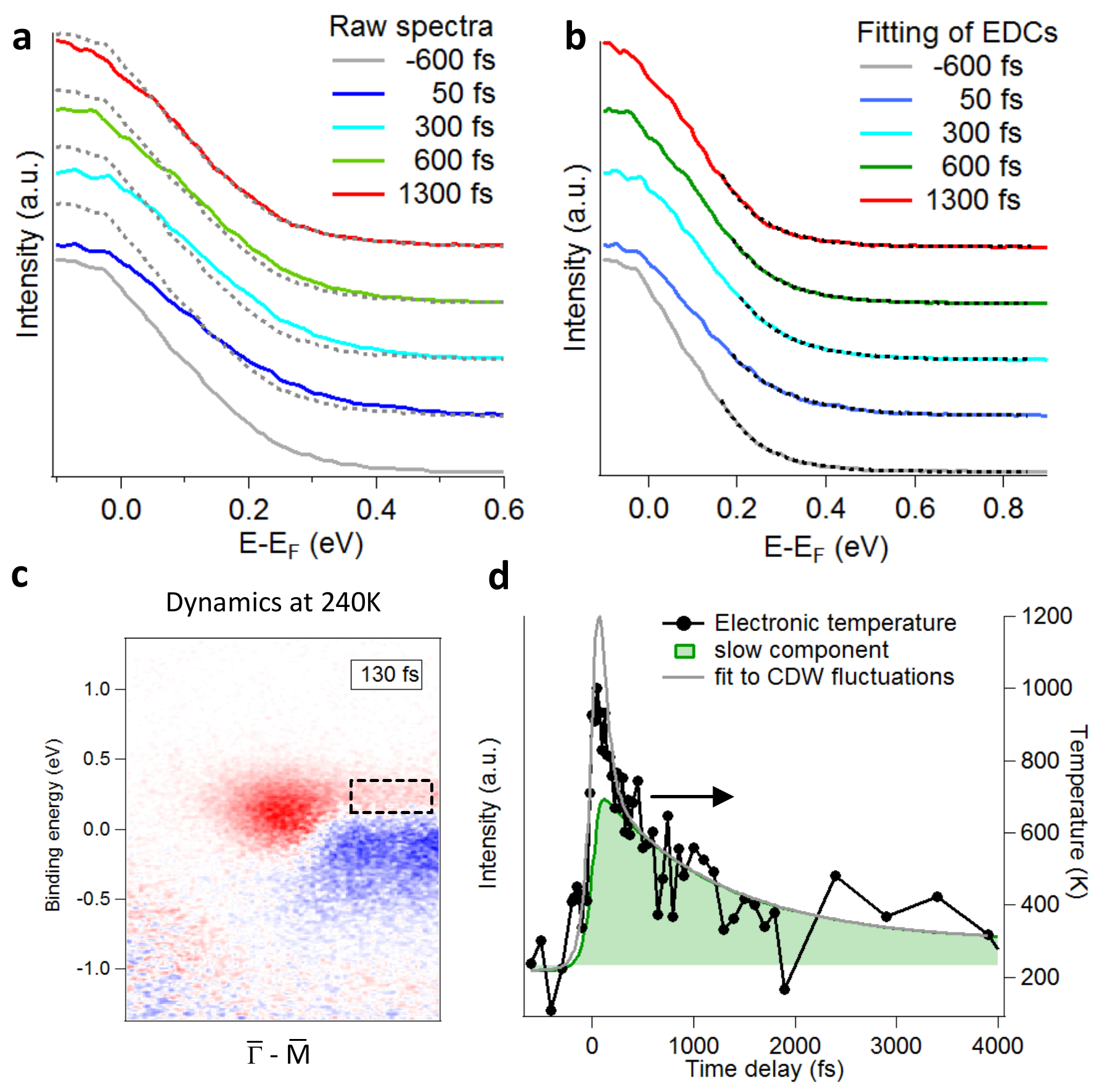}
\end{center}
\caption{\label{fig_5}
(a) EDCs taken at $\bar{M}$ at 240 K (integrated inside the dashed box of graph (c)) for different time delays. The dashed grey line is the reference EDC at -600 fs. (b) The same EDCs fitted by an exponential. (c) False colour plot of the transient photoemission intensity at $\bar{M}$ (same as in Fig. \ref{fig_4}(c)). (d) Comparison of the electronic temperature with the fitted transient photoemisson intensity at the bottom of the CB at $\bar{M}$ (black box in the blue region of Fig. \ref{fig_4} (c)) and the corresponding slow component fitted from the spectral feature (see main text).
}
\end{figure}

We now focus on the dynamics measured in the CB at $\bar{M}$ near the Fermi edge. We exploit here the fact that at this temperature, the CB in our sample is partially filled \cite{MonneyTPRB} and analyse the slope of the corresponding Fermi-Dirac cut-off to monitor the transient electronic temperature, in a similar way to Perfetti {\it et al.} \cite{PerfettiTTM}. Indeed, at $\bar{M}$, at 240 K, there is a flat blue-red region in the difference plot of Fig. \ref{fig_5} (c) (see dashed box), typical from a heated metallic Fermi-Dirac edge.
The (normalized) EDCs in this box are averaged over momentum and plotted in Fig. \ref{fig_5} (a) at different time delays, showing a clear increase of temperature. They are fitted by an exponential function\cite{PerfettiTTM} above $E_F$, in order to avoid the changes of the spectral lineshape at lower binding energies (coming from the photoinduced changes in the spectral function). This is shown in Fig. \ref{fig_5} (b). The extracted electronic temperature (black curve with discs) is displayed in Fig. \ref{fig_5} (d) (after removing the contribution of the energy resolution\cite{GreberTemp} about 300 meV). 
In the spirit of a two-temperature model, the electronic temperature decreases here in about 1 ps as the result of coupling to the lattice, the temperature of which should increase by about 40 K (if heat diffusion is neglected) for the pump fluence used here and the specific heat \cite{CravenHeat} of \tise.

\section{Discussion}
\label{sec_disc}

The reduction of photoemission intensity observed at the bottom of the CB in Fig. \ref{fig_4} (c) comes from a photoinduced change of its spectral function. We interpret this effect as the suppression of the high binding energy tail of the CB, which is the spectral signature of the CDW fluctuations\cite{MonneyPRB12}. This means that the intensity of the CDW fluctuations are reduced by the pump pulse. In Fig. \ref{fig_4} (e), we focus on the dynamics of these CDW fluctuations. The recovery dynamics are fitted with a bi-exponential function (convoluted with a gaussian for experimental time resolution), involving a fast ($\tau\cong 70$ fs) and a slow ($\tau\cong 1220$ fs) component.

To understand these two timescales, we plot in Fig. \ref{fig_5} (d) the fitted time-dependent photoemission intensity at the bottom of the CB (grey line) together with the fitted slow component (green) and the extracted electronic temperature (black ). This comparison suggests that the two timescales of the dynamics of the CDW fluctuations do not match the decay behavior of the electronic temperature (especially the sharp peak near time 0). However, the slow component (green) timescale matches rather well the long timescale of the electronic temperature, indicating that part of the quenching of the CDW fluctuations is due to the transiently increased electronic temperature.

We interpret the fast component of the suppression of the CDW fluctuations as a consequence of the ultrafast screening of the Coulomb interaction, which is enhanced by the new charge carriers created by the optical excitation\cite{RohwerNature}. The intensity of these fluctuations depends strongly on the strength of the interband Coulomb interaction (see Supplementary Materials for an analytical formula of the self-energy of the CB\cite{SuppMat}). In the framework of the excitonic insulator phase, this is the attractive interaction binding an electron and a hole in an exciton. In a previous study\cite{MonneyPRB12}, the resulting self-energy leading to the broadened CB was derived and shown to have a particularly high value at the bottom of the CB. This reveals then that electrons at the bottom of the CB experience a high scattering rate\cite{SentefSelf}, naturally explaining the ultrafast recovery of the CDW fluctuations.

In Fig. \ref{fig_4} (f), we depict how this electron-hole interaction between the valence and conduction bands occurs. Here an electron jumps from the CB to the VB, scattering another electron from the VB to the CB. This can effectively been seen as the scattering of an electron in the CB with a hole in the VB. The strength of the CDW fluctuations depends strongly on the Coulomb interaction and also on the electronic and lattice temperature\cite{MonneyPRB12}. Therefore they tend to be suppressed not only by enhanced screening, but also by the increased transient electronic temperature. 

The combination of these two observations explains the bi-exponential decay of the photoemission intensity suppression at the bottom of the CB (see Fig. \ref{fig_4} (e)), namely the recovery of the CDW fluctuations. We ascribe the fast component ($\tau\cong 70$ fs) to the relaxation of the enhanced screening by the efficient electron-hole scattering and the slow component to a quasi-thermal recovery following the relaxation of the electronic temperature.
Our interpretation that the fast component is due to the electron-hole correlations as a strong scattering channel supports their occurence at temperature above but near $T_c$. These correlations give rise to the CDW fluctuations in this temperature range\cite{MonneyPRB12} and play an important role in stabilizing the low-temperature CDW phase.

At 30 K, the situation is different as a well-formed PLD is present. The longer timescale dynamics of the CDW band display a clear response of the lattice degrees of freedom.  The $A_{1g}$ CDW amplitude mode is observed as a coherent phonon oscillation, provided that the CDW band is selectively excited by a resonant MIR photon pulse generating excited electrons with a minimal excess kinetic energy with respect to the CDW gap.
The fact that the intensity of the CDW band oscillates with the frequency of the $A_{1g}$ CDW-related phonon mode implies that the order parameter of the CDW phase is strongly coupled to the corresponding atomic motion. Furthermore, this coupling is very localized in momentum space, namely only for states in the CDW band. This means that both the excitonic interband Coulomb interaction and the electron-phonon coupling contribute to the order parameter of the CDW phase \cite{MonneyPRB12,KanekoEIphon,vanWezelPRB,PorerNatureMat}. In the first 100 fs, the melting of the CDW order is dominated by the suppression of the excitonic part, as observed by Rohwer {\it et al.} \cite{RohwerNature} with rise times as fast as 20 fs for large fluences (up to 5 mJ/cm$^2$). 

In that sense, our experimental results support a model according to which strong electron-hole correlations, giving rise to short-range CDW fluctuations above $T_c$, contribute to the stabilization of the CDW phase together with the electron-phonon coupling below $T_c$\cite{MonneyNJP,MonneyPRB12}.

\section{Conclusion}
\label{sec_conclusion}

In conclusion, we have probed both the femtosecond dynamics of the CDW phase in \tise\ at 30 K, as well as the dynamics of the high-temperature fluctuation regime at 240 K using trARPES. Below $T_c$, in the CDW phase, we observe the coherent oscillation of the CDW amplitude mode for the first time with trARPES using a mid-infrared excitation resonant with the CDW gap. Above $T_c$, we single out a fast component with a relaxation time $\leq$ 100 fs. We interpret it as the consequence of the ultrafast screening of the interband Coulomb interaction responsible for the electron-hole correlations. We propose that these electron-hole correlations act as an efficient scattering channel in the relaxation of excited electrons. In this way, we have unveiled the dynamics of both the electron and lattice degrees of freedom in \tise.

\section{Acknowledgements}

The authors thank H. Beck, P. Aebi, K. Rossnagel for useful discussions. C.M. acknowledges support by the Swiss National Science Foundation under grant number $PZ00P2\_ 154867$, as well as by the Alexander von Humboldt Foundation. Access to Artemis was provided by STFC.

\end{document}